\documentclass[
amsmath,amssymb]
{revtex4}
\usepackage{pstricks}
\usepackage{graphicx}% Include figure files
\usepackage{dcolumn}% Align table columns on decimal point
\usepackage{bm}% bold math

\begin{document}

\newcommand{\locsection}[1]{\setcounter{equation}{0}\section{#1}}
\renewcommand{\theequation}{\thesection.\arabic{equation}}

\def\F{{\bf F}}
\def\A{{\bf A}}
\def\J{{\bf J}}
\def\af{{\bf \alpha}}
\def\beqn{\begin{eqnarray}}
\def\eeqn{\end{eqnarray}}

\def\dspace{\baselineskip = .30in}
\def\beq{\begin{equation}}
\def\eeq{\end{equation}}
\def\bw{\begin{widetext}}
\def\ew{\end{widetext}}
\def\pl{\partial}
\def\na{\nabla}
\def\al{\alpha}
\def\bt{\beta}
\def\Ga{\Gamma}
\def\ga{\gamma}
\def\de{\delta}
\def\De{\Delta}
\def\da{\dagger}
\def\ka{\kappa}
\def\si{\sigma}
\def\Si{\Sigma}
\def\te{\theta}
\def\La{\Lambda}
\def\lam{\lambda}
\def\Om{\Omega}
\def\om{\omega}
\def\ep{\epsilon}
\def\non{\nonumber}
\def\sq{\sqrt}
\def\sqg{\sqrt{G}}
\def\sp{\supset}
\def\sb{\subset}
\def\l{\left (}
\def\r{\right )}
\def\lq{\left [}
\def\rq{\right ]}
\def\fr{\frac}
\def\la{\label}
\def\hs{\hspace}
\def\vs{\vspace}
\def\inf{\infty}
\def\ran{\rangle}
\def\lan{\langle}
\def\ov{\overline}
\def\tl{\tilde}
\def\tm{\times}
\def\lrar{\leftrightarrow}

%\begin{document}

%\preprint{HD-THEP-08-08}

%\preprint{OSU-HEP-08-01}

%\preprint{February 6, 2008}

\vs{1cm}

\title{Flavor Symmetry and Grand Unification}
% Force line breaks with \\

\author{Berthold Stech}
\email{B.Stech@ThPhys.Uni-Heidelberg.DE}

\affiliation{Institut f\"ur Theoretische Physik, Philosophenweg 16, D-69120 Heidelberg, Germany}
%\author{Zurab Tavartkiladze}%
% \email{zurab.tavartkiladze@okstate.edu}

%\affiliation{Department of Physics, Oklahoma State University, Stillwater, OK 74078, USA}

\vs{1cm}

\date{\today}% It is always \today, today,

\begin{abstract}

The combination of flavor symmetries with grand unification is considered: GUT  $ \times$ flavor . To accommodate three generations the flavor group $SO(3)$ is used. All fermions transform as  3-vectors under this group.  The Yukawa couplings are obtained from vacuum expectation values of flavon fields. For the flavon fields (singlets with respect to the GUT group) and the Higgs fields (singlets with respect to the generation group) a simple form for the effective potentials is postulated. It automatically leads to spontaneous symmetry breaking for these scalar fields. Discrete $S_4$ transformations relate the different  locations of the minima of the potentials.These potentials can be used to describe the hierarchy of the well known up quark mass spectrum. Also the huge hierarchy of the masses of the Higgs fields in grand unified models can be parametrized in this way. It leads to a prediction of the mass of the lightest Higgs boson in terms of its vacuum expectation value $v_0$:  $m_{Higgs} = \frac{v_0 }{ \sqrt{2}} = 123~ GeV$.

\end{abstract}

%\pacs{11.30.Hv, 12.10.Dm, 12.15.Ff, 14.60.Pq}
\maketitle

\section{Introduction}\label{sec:1}

 Grand unified theories \cite{Pati:1974yy} (GUT) provide a clear understanding of the structure and the quantum numbers of the standard model. Because of the existence of 3 generations any GUT symmetry should also be extented by a flavor symmetry.  The simplest extention
uses the direct product  GUT$\times$ Flavor.\\

 There are three major flavor puzzles: i) The extreme smallness of the standard model Higgs and fermion masses compared to the grand unification scale or the mass scale of the heavy neutrinos and other very heavy states. ii) The hierarchy of the fermions themselves as is manifest from the very different values of the up quark masses. iii) the mixing parameters observed for quarks and neutrinos and their difference.\\
 
~~~~         In the literature there are many suggestions for a solution of these puzzles, in particular supersymmetry combined with flavor symmetry or the use of extra dimensions with the idea of different wave functions on the bulk.  However, the results of most models presented in the literature can not be applied here because of their use of different representations of the flavor group in the quark and lepton sectors or for particles and antiparticles: In GUT's considered here, $SO(10)$ and $E_6$, the fermion fields  belong to a single irreducible representation of the GUT group. Thus, all fermions have to belong to a unique representation of the flavor group. For our purpose this also excludes  models which, for instance, have different Frogatt-Nielson charges for up and down quarks. Only few attempts are based on the combination of GUT and flavor symmetries, notably \cite{Medeiros},\cite{Morisi},\cite{Grimus},\cite{Morisi2}. In these papers flavor quantum numbers are assigned to Higgs fields as well as to the fermions. In \cite{ZB}, on the other hand, the Higgs fields are taken to be flavor singlets and particle mixing arises directly from an antisymmetric flavon field combined with an antisymmetric Higgs field. The latter also provides for the tiny mixings of the standard model fermions with high mass states. 

In all these models  the scalar fields (Higgs and flavon fields) are least understood. The reason is that scalar fields are strongly influenced by the structure of the vacuum or, possibly, by its bound state character. So far,  no complete understanding of the scalar sector is in sight and no invariant potentials causing the required symmetry breaking could  be given. In this situation it may be worthwhile to have a phenomenological form  - an  effective potential - which automatically leads to minima at positions which one can easily fix. These effective potentials should allow to describe even very large hierarchies like the one occurring between the vacuum expectation value of the standard model Higgs and its high mass partner in a GUT. 

In this article it is shown, that a very simple form of potentials are suited for this purpose. These effective potentials are fully invariant and need only few parameters which have to be tuned.   Applications important for the three flavor puzzles mentioned above are given.

~~~~In section 2  the spontaneous breaking of the flavor group $SO(3)$ is treated by starting from a flavon field which is symmetric in flavor indices. A flavor invariant potential is constructed. With appropriate parameters its minima provide for spontaneous symmetry breaking corresponding to the well known hierarchy of the up quark masses.\\

~~~~In section 3 we add a flavon field which is antisymmetric in flavor indices.The potential obtained is suited for the vacuum expectation values wanted for a quantitative description of  quark and neutrino mixings in the $E_6$ GUT model \cite{ZB} . \\ 

~~~~~In section 4 we deal with the $SU(3)_L \times SU(3)_R$ Higgs field needed in  $E_6$ \cite{ZB} which appears at the scale of the electro weak unification. Its minima should simultaneously give the scale of the top quark, the bottom quark and the high mass of the heavy down quark type fermion (related to the heavy "right handed" neutrinos). It is shown that these masses can be described by the suggested form of the potential. At the same time one also gets the huge splitting between the low and high masses of the Higgs field and even a prediction for the mass of the standard model Higgs. \\

\section{ Flavor symmetry breaking and the up quark hierarchy.}\label{sec:2}

 To accommodate 3 generations we use the flavor group $ SO(3)$. All fermions transform as vectors with respect to this group. The product of two fermions  in the Yukawa interaction then transform as $"1" + "5" $ representations for a symmetric combination and as a $"3"$ for the antisymmetric combination. To obtain an invariant interaction there is then the possibility to give all Higgs fields generation indices. But this generally increases their numbers drastically. Instead, one can interpret the  coupling matrices multiplying the Higgs fields as vacuum expectation values of new scalar fields (flavons), which are GUT singlets but carry the necessary generation quantum numbers. We choose here the second alternative and thus keep the Higgs fields to be singlets in generation space.

Using left handed two component Weyl fields $\psi^\alpha$ for the fermions (with $\alpha = 1,2,3 $ denoting the generations), the Yukawa interaction is of the form 

\beq
\la{Yu}
  {\cal L}^{eff}_Y  =   \frac{\langle \Phi_{\alpha \beta} \rangle}{M} (\psi^{\alpha T} H \psi^{\beta})+..\\
 \eeq
     
  $\Phi_{\alpha \beta}$ describes real flavon fields, $  H $ a Higgs field, and $M$ gives the scale at which the effective Yukawa interaction of dimension 5 is formed. In (\ref{Yu}) GUT indices are suppressed.

 %\begin{eqnarray}
%\label{1.1}
%\Phi(x) = \chi(x) + {\rm i} \xi(x)~.
%\end{eqnarray}
Clearly, Yukawa interactions of this form are effective ones and have to be understood on a deeper level. However, in this article I will be concerned with the phenomenology of the effective Yukawa interaction only. The lowest Higgs representations ($"10"$ in $SO(10) $~and $"27"$ in $E_6$) are symmetric representations. Thus, for these Higgs fields the Pauli principle requires a symmetric representation for the flavon fields connected with them, i.e.  $"1"$ and $"5"$ representations with respect to $SO(3)$. We describe this part of the flavon field $\Phi$ by a real and symmetric $3 \times 3 $ matrix $\chi_{\alpha,\beta}$. By an orthogonal transformation, which can be absorbed by the fermion fields, this matrix can be taken to be diagonal as in \cite{ZB}. This choice defines a direction in symmetry space for a possible spontaneous symmetry breaking.

\begin{eqnarray}
\label{diag}
 \Phi_{\alpha \beta} 
~~~~~~~~~~~~~~~~\Rightarrow  
\hspace{1cm}
\begin{array}{ccc}
& {\begin{array}{ccc}
 & &
\end{array}}\\ \vspace{2mm}
~~~~~~~~ \hs{-0.5cm}
\begin{array}{c}
  \\
\end{array} \hspace{-0.1cm}&{\left(\begin{array}{ccc}
\chi_1 & 0& 0
\\
0 & \chi_2 & 0
\\
0 & 0 & \chi_3
\end{array}\right)~.
}
\end{array}
\end{eqnarray}
There  still remains the freedom of such $ SO(3) $ transformations which keep $ \chi $ diagonal. This remaining symmetry is necessarily a discrete subgroup of $ SO(3)$.  It is the group $S_4$. This group simply permutes the $\chi$ fields together with the fermion generations.  $S_4$  has been suggested as a symmetry or intermediate symmetry in many publications starting with \cite{S4}, discussing $S_4$ invariant Higgs potentials \cite{Lin} etc. In our treatment  $S_4$  plays a different role. As we will see, it will not occur as an intermediate symmetry which is finally broken. The proposed potential will fully break $SO(3)$ in one step. But obviously, because of the complete $SO(3)$ invariance of the potential which we have to construct, each $S_4$ permutation of the $\chi $ values at the minimum of the potential will also be a minimum.  

The part of the Yukawa interaction (\ref{Yu}) with (\ref{diag}) does not  induce particle mixing. Instead, it determines the hierarchy of the fermion masses. The effective potential we are looking for should describe the mass spectrum of the up quarks. (The spectrum of the down quarks and charged leptons is closely related to the up quark spectrum at least in the $E_6$ model of ref. \cite{ZB}). 
Let us then consider three   $SO(3)$ invariants formed from the matrix $\chi_{\alpha,\beta} $

\beq
\label{inv}
 J_1= ( Tr[\chi])^2,\qquad J_2= Tr[\chi \cdot \chi],\qquad J_3= Tr[\chi \cdot \chi \cdot \chi \cdot \chi]~.
 \eeq
 
 These invariants can now be used to form the effective potential suggested here:

\beq
\label{V1} 
 V(\chi) = c_1  M^2 J_1~ (\log{\frac{J_1}{\mu_1^2}} -1) + c_2 M^2 J_2 ~(\log{\frac{J_2}{ \mu_2^2}}-1) +
 c_3 J_3~(\log{\frac{J_3}{\mu_3^4}}-1)~.
 \eeq
 
 $M$ describes the scale of the field $\chi$. To get simple expressions for the final results the numbers "-1" are not incorporated into the log terms. Now the values of $\mu_1$, $\mu_2$ and $\mu_3$  fix the minima $\lan \chi\ran$ of the potential. The coefficients $c_1, c_2, c_3$  are not relevant for the positions of the minima but have to be non vanishing positive numbers. They affect the strength of the second derivatives of $V(\chi)$. These second derivatives of $V$ form a $ 3 \times 3$ matrix which is positive definite at $ \chi = \lan \chi\ran$ (for properly chosen signs of the invariants) .
 
 Let us require that a minimum occurs at $\lan\chi_1\ran/M= m_u/m_t=\sigma^4 $, $\lan\chi_2\ran/M= m_c/m_t =\sigma^2 $ and $\lan\chi_3\ran/M=1$ with $\sigma=0.050$ which pretty well describes the hierarchy of the up quark masses \cite{ZB}. The values of the parameters $\mu$ in (\ref{V1}) are  easily  obtained by requiring
\beq
\label{V'}
\frac{\partial V}{\partial \chi_1 }= 0, \qquad \frac{\partial V}{\partial \chi_2} = 0,\qquad \frac{\partial V}{\partial \chi_3} = 0
\eeq
taken at the above values for the $\lan \chi\ran$'s. One finds:
\beq
\label{mu}
\mu_1^2 = \langle J_1 \rangle =( 1+\sigma^2+\sigma^4)^2  ~M^2, \qquad \mu_2^2 = \langle J_2 \rangle = (1+\sigma^4+\sigma^8)~ M^2,\qquad  \mu_3^4 = \langle J_3 \rangle = (1+\sigma^8+\sigma^{16}) ~M^4 .
\eeq
 
 Here $ \langle J_i \rangle $ denotes the value of $J_i$ at the designed minimum.
Thus, the potential obtained has a minimum at the required position. Spontaneous symmetry breaking is induced. The same minimum appears at positions obtained by permutations of the three $\chi$ values according to the $S_4$ symmetry in our diagonal basis.  \\

The eigenvalues of the matrix for the second derivatives at the minimum give the square of the masses of the flavon fields. They depend on the coefficients $c$ for which no reliable theory is available. But because of the importance of the magnitude of the flavon masses a meaningful suggestion may be useful:
Let us consider the three different potentials occurring in (\ref{V1}) separately. The second derivatives at the minimum of the $i^{th}$ potential forms a $3\times 3$  matrix which will be denoted by $c_i L_i$.  It has two zero mass eigenvalues and one non vanishing eigenvalue. The latter can directly be obtained from the trace of $ c_i L_i$ and is equal to the square of the mass of the flavon field in case no other potential term is present. It is then suggestive to identify this term with the mass scale $\mu_i$. This way one can find the coefficient $c_i$ which "normalizes" the $i^{th}$  potential.
\beq
\label{TrL}
 c_i = \frac{\mu_i^2}{Tr[L_i] }.
 \eeq 

The full potential is now taken to be the sum of these "normalized" potentials with equal weights. This strong assumption is clearly highly speculative but worth trying. It is a kind of a "bootstrap" condition for the coefficients occurring in (\ref{V1}).\\
 
 From(\ref{TrL}) one gets
\beq
\label{fixc}
c_1 = \frac{1+s^2+s^4}{12},\qquad c_2 = \frac{1+s^4+s^8}{4},\qquad c_3 = \frac{1}{16}~ \frac{(1+s^{8}+s^{16})^{3/2}}{1+s^{12}+s^{24}}. 
\eeq

By adopting these coefficients one finds for the flavon masses from (\ref{V1}) and (\ref{mu})
\beq
\label{NM}
\frac{M_1}{M} = 1.57, \qquad \frac{M_2}{M} =0.736 ,\qquad\frac{M_3}{M} = 0.00125.
\eeq

The strength of $V$ at the minimum is $ -0.396~M^4$. 

\section{ Flavor symmetry and mixings.}\label{sec:3}

The flavon fields $\Phi$ coupled to the fermions in the effective Yukawa interaction will also have a part antisymmetric with respect to $ SO(3)$ flavor indices. As mentioned above it is a $"3"$ of $SO(3)$. It can only go together with the antisymmetric Higgs representations  $"120"$ in $SO(10)$ and $"351"$ in $E_6$. It clearly leads to generation mixing. \\$E_6$ has the advantage that generation mixing can be combined with the mixing of standard model particles with heavy states, necessary in all GUT models.

The corresponding flavon field is described by the antisymmetric $ 3 \times 3$ matrix field  $\xi_{\alpha, \beta}$ which we take to be hermitian:

\begin{equation}
\label{xi}
\xi:  =
 {\rm i}~\left(
\begin{array}{ccc}
0&\xi_3&-\xi_2\\
-\xi_3&0&\xi_1\\
\xi_2&-\xi_1&0
\end{array}\right)~.
\end{equation}

The potential to be constructed should lead to vacuum expectation values for the fields in (\ref{xi}) in the basis in which the matrix $\chi$ is diagonal. In this basis, one has again the discret symmetry $S_4$ which simultaneously permutes the entries in $\xi$ as well as those in $\chi$. In order to obtain a potential having a minimum which fixes all 6 flavon fields remaining in this basis one needs 3 additional invariants. They are chosen to be
\beq
\label{inv2}
  J_4=  Tr[\xi \cdot \xi],\qquad J_5= Tr[\xi \cdot \chi \cdot \xi \cdot \chi],\qquad J_6 =Tr[\xi \cdot \xi \cdot \chi \cdot \chi].
\eeq

For the total potential we simply add the corresponding terms:

\beq
\label{Vt}
V( \chi,\xi) =  V(\chi) + c_4 M'^2 J_4 ~ (\log{\frac{J_4}{ \mu_4^2}}-1) +c_5 J_5~ (\log{\frac{J_5}{\mu_5^4}}-1)+
c_6 J_6 ~(\log{\frac{J_6}{\mu_6^4}}-1)~.
\eeq

As in section 2, it is easy to force this potential to have a minimum at prescribed values by simply putting all first derivatives equal to zero at the required positions $\langle\chi \rangle $ and 
$\langle \xi \rangle $. 
One obtains 
\beq
\label{mu456}
 \mu_4^2 = \langle J_4 \rangle, \qquad \mu_5^4 =  \langle J_5 \rangle, \qquad \mu_6^4 =  \langle J_6 \rangle. 
 \eeq 
Thus one can choose vacuum expectation values for $\chi$ and $\xi$ which can be used in GUT models in order to describe simultaneously the fermion hierarchy and the fermion mixings.
Again, the same minimum of the potential occurs for simultaneous $S_4$  transformations 
of $\langle \chi \rangle $ and $\langle \xi \rangle $ . Notably,
the $V(\chi)$ part of the total potential remains unchanged. However, the matrix for the 
second derivatives is now a $ 6 \times 6 $ matrix with  6 positive eigenvalues. \\

To be able to calculate the 6 boson masses one needs besides (\ref{fixc}) the coefficients $c_4, c_5, c_6.$  
 Based on the speculation mentioned in section 2 one may use again (\ref{TrL}) and then obtains
 \beq
\label{c456}
 c_4 = \frac{\langle \xi_1\rangle^2+\langle \xi_2\rangle^2+\langle \xi_3\rangle^2}{4 M'^2},
 \qquad c_5 \simeq \frac{s^3 \langle \xi_1 \rangle^3 M}{\sqrt{2}~(\langle \xi_1\rangle^4+\langle \xi_2\rangle^4)},\qquad c_6 \simeq \frac{\sqrt{\langle \xi_1\rangle^2+\langle \xi_2\rangle^2} M }{4~(
 M^2+\langle \xi_1\rangle^2+\langle \xi_2 \rangle^2 )}
 \eeq
For simplicity  $c_5$ and $c_6$ are approximated by taking only the smallest power in $s$. The potential (\ref{Vt}) is now fixed.\\

In \cite{ZB} - in the framework of an $E_6$ model - values for $\langle\chi \rangle $ and 
$\langle \xi \rangle $ are used together with only a few more parameters for a quantitative fit for the masses, mixings and $CP$ properties of all fermions. In this  $E_6$ GUT a Higgs field in the "27" representation of $E_6$ connects low and high scales. In the next section the potential approach is applied to the important $SU(3)_L \times SU(3)_R$ part of this field.

\section{ Higgs Fields with low and high scale vacuum expectation values.}\label{sec:4}

An interesting example of a scalar Higgs field in a GUT is the field $ H_{27} $, the irreducible "27" representation of $ E_6$. In \cite{ZB} the breaking of $E_6$ leads to the intermediate symmetry $ SU(3)_L \times SU(3)_R \times SU(3)_C$. It covers the region from $\approx 2 \cdot 10^{13}~ GeV $ (the point of electro weak unification) up to the complete gauge group unification at $\approx 10^{17}~ GeV$. The vacuum expectation values of $ H_{27} $ necessarily occur in the  $ SU(3)_L \times SU(3)_R $ part. Thus, we can study the corresponding $ 3 \times 3 $ matrix field $ H^i_k $ where the index $i$ transforms as an upper index with respect to $ SU(3)_L $ while the lower index $k$ transforms according to $ SU(3)_R$. The indices $i=1,2$ are the $SU(2)_L$ indices of the standard model.

By absorbing transformations and phases by the fermion fields one can choose a basis in which $ H^i_k $ is diagonal and contains only real and positive elements:
\begin{eqnarray}
 H^i_k 
~~~~~~~~~~~~~~~~\Rightarrow  
\hspace{1cm}
\begin{array}{ccc}
& {\begin{array}{ccc}
 & &
\end{array}}\\ \vspace{2mm}
~~~~~~~~ \hs{-0.5cm}
\begin{array}{c}
  \\
\end{array} \hspace{-0.1cm}&{\left(\begin{array}{ccc}
H_1 & 0& 0
\\
0 & H_2 & 0
\\
0 & 0 & H_3
\end{array}\right)~.
}
\end{array}
\end{eqnarray}

 $H_1$ couples to the top quark, $H_2$ to the bottom quark and $H_3$ to the heavy standard model singlet down quark state $D$. Furthermore, the vacuum expectation value of $H_1$ is to be identified with the vacuum expectation value of the standard model Higgs:  ~$\langle H_1 \rangle = v_0 =174~ GeV$. Thus, the $ SU(3)_L \times SU(3)_R $ Higgs field $H$ should have the vacuum expectation values (see \cite{ZB})

\beq
\label{HVeV}
\langle H_1\rangle = v_0,\qquad \langle H_2\rangle \approx m_b, \qquad \langle H_3 \rangle \approx m_D \approx 2 \cdot 10^{13} ~GeV~.
\eeq

By our suggested form of effective potentials this can easily be achieved.
Defining the  $ SU(3)_L \times SU(3)_R $ invariants
\beq
\label{Y}
Y_1 = Tr[H \cdot H^\dagger], \qquad Y_2 = \det{H}, \qquad Y_3= Tr[ H\cdot H^\dagger \cdot H \cdot H^\dagger],
\eeq

the potential is taken to be

\beq
\label{VY}
V(H) =c_1 M^2 Y_1~(\log{\frac{Y_1}{\mu_1^2}} - 1) +  c_2 M Y_2~ (\log{\frac{Y_2}{\mu_2^3}} - 1) + 
 c_3 Y_3~ (\log{\frac{Y_3}{\mu_3^4}} - 1) .
 \eeq

As in the cases discussed before, the physics input (\ref{HVeV})  fixes the potential apart from the coefficients $c$.

\beq
\label{muH}
\mu_1^2 = \langle Y_1 \rangle = m_t^2 +m_b^2 +m_D^2 , \qquad \mu_2^3 = \langle Y_2 \rangle  = m_t m_b m_D,\qquad 
\mu_3^4 = \langle Y_3 \rangle = m_t^4 + m_b^4 + m_D^4 .
\eeq

Thus, even a huge hierarchy can be accommodated: $v_0 = 174 ~GeV$, $m_b = 2.9~ GeV$  and $m_D =M= 2 \cdot 10^{13}~GeV$  (all taken at the scale of the $Z$ boson).  The subgroup $S_4$ of $SU(3)_L \times SU(3)_R $ 
permutes the three $\langle H_i \rangle$ values without changing the minimum. \\

For calculating the second derivatives of the potential and the three eigenvalues of the 
corresponding $3 \times  3$ matrix one has to fix the new coefficients $c_1, c_2, c_3 $ for the present case. As in section 2 and 3 it is suggestive to apply (\ref{TrL}). One obtains this way:

\beq
    c_1=\frac{m_b^2+m_t^2+M^2}{4 M^2}, \qquad  c_2 = \frac{(m_b m_t )^{5/3} M^{2/3}}{m_b^2 m_t^2 +m_b^2 M^2  +m_t^2 M^2},\qquad c_3=\frac{1}{16}
     \frac{(m_b^4+m_t^4+M^4)^{3/2}}{m_b^6+m_t^6+M^6}.
 \eeq
 
With these coefficients the spontaneous symmetry breaking of the $ SU(3)_L\times SU(3)_R $ Higgs fields  leads to the masses
\beq 
\label{HM}
M_1 (\langle H \rangle) =  2.83 \cdot10^{13}~ GeV ,\qquad M_2 (\langle H \rangle) = 2.15 \cdot 10^{5} ~GeV , \qquad M_3 
(\langle H \rangle) =  123~ GeV. 
\eeq

The value of $V$ at the minimum turns out to be $- 0.313~M^4$.\\
The most uncertain mass in (\ref{HM}) is $M_2$. The reason is that the value $\mu_2$ which determines $c_2$  differs strongly from $\mu_1$ and $\mu_3$. For instance, if one would replace in (\ref{TrL}) $\mu_i$ by the general mass scale $M$ there would be no change of the numerical results given in
(\ref{NM}), and no change for $M_1$ and $M_3$ in (\ref{HM}). But $M_2$ would get a very high mass value. Fortunately and remarkably,  the interesting light Higgs mass $M_3$ is insensitive to changes of scales: \\
The value  $M_3 $ is independent of $M$ for large $M$ and independent 
of $m_b$ for $m_b << v_0$. One finds
\beq
M_3  = m_{Higgs} = \frac{v_0}{\sqrt{2} } = 123~GeV.
\eeq
\\
The two assumptions (logarithmic potentials and "bootstrap" determination of coefficient factors) are predictive but  speculative.  Besides, quadratic divergences can fully ruin the picture given here. However, quadratic divergencies have their origin in tadpole graphs and tadpole contributions can be subtracted by momentum subtraction or by other means. Thus, there is a chance that the effective potentials for scalar fields suggested here is useful at least for 
phenomenological studies.\\

\newpage

{\bf Acknowledgment}

\vs{0.2cm}
I like to thank Dieter Gromes and Matthias Neubert for valuable discussions.

\bibliographystyle{unsrt}

\begin{thebibliography}{99}


%\cite{Pati:1974yy}
\bibitem{Pati:1974yy}
  J.~C.~Pati and A.~Salam,
  %``Lepton Number As The Fourth Color,''
  Phys.\ Rev.\  D {\bf 10} (1974) 275;
H.~Georgi and S.~L.~Glashow,
  %``Unity Of All Elementary Particle Forces,''
  Phys.\ Rev.\ Lett.\  {\bf 32} (1974) 438.

%\bibitem{E6}
%F.~Gursey, P.~Ramond and P.~Sikivie,
%  %``A Universal Gauge Theory Model Based On E6,''
%  Phys.\ Lett.\  B {\bf 60} (1976) 177;
%Y.~Achiman and B.~Stech,
%  %``Quark Lepton Symmetry And Mass Scales In An E6 Unified Gauge Model,''
%  Phys.\ Lett.\  B {\bf 77} (1978) 389;
%Q.~Shafi,
%  %``E(6) As A Unifying Gauge Symmetry,''
%  Phys.\ Lett.\  B {\bf 79} (1978) 301;
%R.~Barbieri, D.~V.~Nanopoulos and A.~Masiero,
%  %``Hierarchical Fermion Masses In E6,''
%  Phys.\ Lett.\  B {\bf 104} (1981) 194.
\bibitem{Medeiros}
I. de Medeiros Varziels and G.G, Ross,
 Nucl. Phys. B {\bf 733}, 31 (2006), hep-ph/0507176.

\bibitem{Morisi}
S. Morisi, M.Picariello and E.Torrente-Lujan
Phys. Rev. D{\bf 75}, 075015 (2007), hep-ph/0702034.

\bibitem{Grimus}
W. Grimus and H. Kuhbock,
Phys. Rev. D {\bf 77}, 055008 (2008), hep-ph/0710.1585.

\bibitem{Morisi2}
F. Bazzocchi, M. Frigerio and S. Morisi,
Phys. Rev. D{\bf 78}, 116018 (2008), hep-ph/0809.3573.

\bibitem{ZB}
  B.~Stech and Z.~Tavartkiladze,
  Phys. Rev. D {\bf 77} (2008) 076009,
  hep-ph/0311161].\\
B. Stech,
Fortsch.Phys. {\bf 58} : 692-698 (2010),
hep-ph/1003.0581.

\bibitem{S4}
S. Pakvasa and H. Sugawara 
Phys. Lett. B{\bf 82} (1979) 105.

\bibitem{Lin}
C.~Hagedorn, M.~Lindner and R.~N.~Mohapatra,
  %``S(4) flavor symmetry and fermion masses: Towards a grand unified theory  of
  %flavor,''
  JHEP {\bf 0606} (2006) 042;

%\bibitem{eric}
%For an early review see
%B. Stech, Ettore Majorana Inst. Sci., Vol. 7 (1980) p. 23,
%eds. S. Ferrara et al.

%
%\bibitem{newE6}
%Y.~Achiman  and A.~Lukas,
%  %``Heavy top in an E(6) model for fermionic masses and mixing,''
%  Nucl.\ Phys.\  B {\bf 384} (1992) 78;
%J.~L.~Rosner,
%  %``Splitting between up type and down type quark masses via mixing with exotic
%  %fermions in E6,''
%  Phys.\ Rev.\  D {\bf 61} (2000) 097303;
%J.~D.~Bjorken, S.~Pakvasa and S.~F.~Tuan,
%  %``Yet another extension of the standard model: Oases in the desert?,''
%  Phys.\ Rev.\  D {\bf 66} (2002) 053008;
%%  [hep-ph/0206116];\\
%M.~Bando and T.~Kugo,
%  %``Neutrino masses in E(6) unification,''
%  Prog.\ Theor.\ Phys.\  {\bf 101} (1999) 1313;
% % [arXiv:hep-ph/9902204].
% {\it ibid} {\bf 109} (2003) 87;
%N.~Maekawa and T.~Yamashita,
%  %``E(6) unification, doublet-triplet splitting and anomalous U(1)A
%  %symmetry,''
%  Prog.\ Theor.\ Phys.\  {\bf 107} (2002) 1201;
% % [arXiv:hep-ph/0202050].
%J.~Harada,
%  %``Hypercharge and baryon minus lepton number in E(6),''
%  JHEP {\bf 0304} (2003) 011.
% % [arXiv:hep-ph/0305015].
%  B.~Stech and Z.~Tavartkiladze,
%  %``Fermion masses and coupling unification in E(6): Life in the desert,''
%  Phys.\ Rev.\  D {\bf 70} (2004) 035002.
%%  [hep-ph/0311161].

%
%\bibitem{recentE6}
% C.~R.~Das and L.~Laperashvili,
%  %``Are preons dyons? Naturalness of three generations,''
%  Phys.\ Rev.\  D {\bf 74} (2006) 035007;
% % [hep-ph/0605161];\\
% J.~Sayre, S.~Wiesenfeldt and S.~Willenbrock,
%  %``Dimension-five operators in grand unified theories,''
%  Phys.\ Rev.\  D {\bf 75} (2007) 037702;
%%  [hep-ph/0605293];\\
% F.~Caravaglios and S.~Morisi,
%  %``Gauge boson families in grand unified theories of fermion masses: E_6^4   x
%  %S_4,''
%  Int.\ J.\ Mod.\ Phys.\  A {\bf 22} (2007) 2469.
% % [hep-ph/0611078].
% S.F. King, S. Moretti, R. Nevzrov, Phys.Rev. D {\bf73} (2006) 035009
% 
%\bibitem{S+T:2008sb}
%  B.~Stech and Z.~Tavartkiladze,
%  ``Genertion Symmetry and E6 unification''
%  Phys.\ Rev.\  D {\bf 77} (2008) 076009.
%  [hep-ph/0311161].

%\bibitem{antE6}
%B.~Stech,
%  %``Quark Masses, Charged Current Mixing Angles And CP Violation,''
%  Phys.\ Lett.\  B {\bf 130} (1983) 189;
%B.~Stech,
%Phys.\ Rev.\ D {\bf 62} (2000) 093019.

%\bibitem{Katrin}
%A.Osipowicz et al. [KATRIN Collaboration], arXivhep-ex/0109033.

%\bibitem{Gerda}
%Nucl.Phys.Proc.Suppl. 188:68-70, 2009.

%\bibitem{Fernandez}
%E. Fernandez-Martinez, M.B. Gavela, J. Lopez-Pavon and O. Yasuda,
%Phys. Lett. B {\bf 649}, 427 (2007) [arXiv:hep-ph/0703098]

%\bibitem{Fogli}
%G.L. Fogli, E. Lisi, A. Marrone, A.Palazzo and A.M. Rotunno, Phys. Rev. Lett. {\bf 101},
%141801 (2008).

\end{thebibliography}

 \end{document}